\documentstyle[jeep,graphics,equations,twocolumn,doublespace]{article} 

\begin{document}           % End of preamble and beginning of text. 
\newpage

\begin{center}

{\Large {\bf Dynamics of clade diversification on the morphological
hypercube}}\\  
 
\vspace{.2in} 
 
{\bf Sergey Gavrilets}\\   
 
\end{center} 
 
%\vspace{.25in}  
  
Departments of  Ecology and Evolutionary Biology and  Mathematics, 
University of Tennessee, Knoxville, TN 37996-1610; gavrila@tiem.utk.edu\\ 
 
%\vspace{1in} 
Brief title: Diversification on the hypercube\\ 
 
%\vspace{1in} 

%\newpage 
\section{SUMMARY} 
 
Understanding the relationship between taxonomic and morphological  
changes is important in identifying the reasons for  
accelerated morphological diversification early in the history of  
animal phyla. Here, a simple general model describing the  
joint dynamics of taxonomic diversity and morphological disparity 
is presented and applied to the data on the diversification of blastozoans. 
I show that the observed patterns of deceleration in clade diversification 
can be explicable in terms of the geometric structure of the
morphospace and the effects of extinction and speciation on morphological
disparity without invoking major declines in the size of morphological
transitions or taxonomic turnover rates. 
The model allows testing of hypotheses about patterns of diversification
and estimation of rates of morphological evolution.  In the case of
blastozoans, I find no evidence that major changes in evolutionary rates
and mechanisms are responsible for the deceleration of morphological
diversification seen during the period of this clade's expansion.
At the same time, there is evidence for a moderate decline in overall rates of
morphological diversification concordant with a major change (from
positive to negative values) in the clade's growth rate.

\section{1. INTRODUCTION} 
 
Rapid morphological diversification early in a clade's history, at  
relatively low taxonomic diversity, with an apparent slowdown afterwards 
represents a commonly observed pattern of radiation of animal life. 
Among the best  known examples are Paleozoic blastozoans 
(Foote 1992; Wagner 1995a), bryozoans (Anstey \& Pachut 1995) and gastropods 
(Wagner 1995b), Paleozoic and Mesozoic crinoids (Foote 1994, 1995, 1996b),
Cambrian marine  
arthropods (Briggs {\em et al}. 1992; Wills {\em et al.} 1994) and  
Ordovician trilobites  
(Miller \& Foote 1996). The pattern of rapid initial increase in  
morphological disparity which remains unsurpassed during the history of  
the clade afterwards has often been interpreted as an evidence for  
major secular changes in evolutionary rates and mechanisms (Valentine 1969,  
1980). Different explanations for these  secular changes have been  
proposed. It has been argued that ecological opportunities were greater in  
the early history of many clades, that genetic and developmental systems  
were less canalized early on, and that the nature of adaptation on a  
``rugged'' adaptive landscape results in a slowdown of the rate of adaptation
(e.g., Erwin 1994; Erwin {\em et al.} 1987; McShea 1993; Valentine 1969, 1980; 
Valentine {\em et al}. 1994; Kauffman 1993). Each of these factors  
can potentially 
cause a reduction in the probability and/or size of morphological changes  
with time which will translate into a decline in the rate of clade's  
diversification. 
 
In spite of the well-recognized potential importance of the  
pattern and extensive discussions of its generality and possible explanations, 
there have been only few attempts to use formal mathematical models to
identify 
a minimum set of factors sufficient for explaining the pattern and to test 
hypotheses about underlying mechanisms. 
Existing time-homogeneous models have predicted a linear increase in  
morphological variance (Slatkin 1981; Foote 1991, 1996a; Valentine {\em et
al}. 
1994) reinforcing the belief that something has to change significantly 
in time to result in the observed patterns.  
However, some difficulties arise when one tries to apply these models to
data. One reason is that while the models consider {\em single} traits 
that vary {\em continuously} and whose evolution is {\em unconstrained}, the
empirical studies of morphological disparity are typically based on a 
{\em large number} of {\em discrete} characters which are always subject to 
some morphological {\em constraints} (geometric,  structural, or functional limits on possible trait values). Another reason is that previous
modeling frameworks did not include some of the factors that can
significantly affect the dynamics of morphological disparity (such as 
subclade extinction or origination events that do not result in large
differences between sister-species). Here, I extend the previous work 
by constructing a more detailed model of clade diversification specifically 
designed for treating discrete characters and by applying it to the data 
on the diversification of blastozoans.

\section{2. MODEL} 
 
I consider the evolution of a monophyletic clade driven by extinction, 
speciation and anagenetic changes. Let us assume that each lineage in the  
clade is characterized by $L$ binary morphological traits. A lineage 
$\alpha$ can be described by a sequence of 0's and 1's of length $L$: 
$l^{\alpha}=(l^{\alpha}_1,l^{\alpha}_2,\dots,l^{\alpha}_L)$ where  
$l^{\alpha}_k=0$ or $1$ ($k=1, \dots, L$). The morphological space 
is mathematically equivalent to a binary hypercube. In discussing the 
clade's diversification, it is useful to visualize each lineage as a point 
on a vertex of the morphological hypercube. Accordingly, a clade will be a  
cloud of points. 
Speciation, extinction and anagenesis change the size, location and  
structure of this cloud. Let us define  
a morphological ``distance'' between lineages $\alpha$ and $\beta$  
as the number of traits at which the lineages are different: 
      \begin{equation} \label{dis} 
            d^{\alpha \beta} = \sum_{i=1}^L (l^{\alpha}_i-l^{\beta}_i)^2. 
      \end{equation} 
I will be interested in the joint dynamics of the number of lineages  
in the clade, $N$, and the average pairwise distance within the clade,  
$D=\sum_{i < j} d^{i j}/(N(N-1)/2)$. Distance $D$ is a measure of
morphological  
disparity characterizing the spread of the clade in the morphological space. 

The dynamics of the clade size  
and the morphological disparity have been the focus of the previous work  
(e.g. Slatkin 1981; Foote 1991, 1996a; Valentine {\em et al}. 1994).  
Here in addition to $N$ and $D$, I will consider  
the average distance of the members of the clade from its species-founder,  
$d$. Let $\epsilon$ denote the species-founder of the clade. Then  
$d=\sum_i d^{\epsilon i}/N$. Distance $d$ characterizes the extent of the  
evolution of the clade from its ancestral state. Below this measure will
prove  
to be very informative and convenient to use in analyzing real data. 
I will model clade  
evolution as a random walk on the morphological hypercube with births and  
deaths. That is I will assume that  
origination and extinction events as well as morphological changes can be 
considered as random and independent of the morphology (cf. Raup \& Gould 
1974; Slatkin 1981; McKinney 1990; Foote 1991, 1996a; Valentine {\em et al} 1994).  
This represents a null hypothesis which must be
rejected before introducing additional factors to explain the observed
patterns of taxonomic and morphological diversification. 
 
It is convenient to formulate the model in discrete time.  
I consider two types of morphological changes: anagenetic and  
cladogenetic. Anagenetic changes are modeled by assuming that during a unit  
time interval each trait in a lineage may evolve to an alternative state  
with a small probability  
$\mu_1$. I assume that there are two types of origination events having  
probabilities $\sigma_1$ and $\sigma_2$, respectively. The origination 
events of the first type 
do not result in any immediate differences between two (or more) new 
lineages the old lineage has split into. This might be the case when 
different
large parts of a subdivided populations become completely isolated by  
geographic (as in the vicariance speciation scenario, e.g. Lynch 1989)  
or reproductive (as in the parapatric speciation scenario, e.g.  
Gavrilets 1999; Gavrilets {\em et al}. 1998) factors. The origination events
of  
the second type are accompanied by (significant) morphological changes  
(Eldredge \& Gould 1972). This might be the 
case when speciation takes place in a small (peripheral) population which 
has undertaken significant morphological evolution before emerging as a new 
species (as in the peripatric speciation scenario, e.g. Mayr 1963).  
During such speciation events each trait  
in a new lineage can evolve to an alternative state with probability $\mu_2$. 
I assume that there are two types of extinction events. A 
lineage can become extinct individually (with probability $\delta_1$) or as 
a member of a ``$T$-subclade'' simultaneously with all other members  
(with probability $\delta_2$). Following  Derrida \& Peliti (1991),  
I say that  
two lineages belong to the same $T$-subclade if their last common ancestor  
existed $T$ years ago. Extinction of a subclade might happen when some traits
that are shared by the members of the subclade ``promote''  
extinction (McKinney 1997) (say, after a change in the environment). 
I will assume that all rates defined above are small ($\mu_i,\sigma_i, 
\delta_i <<1, i=1,2$). 
 
The changes in $N,D$ and $d$ between subsequent time intervals are described 
by a system of difference equations (see Appendix for details) 
      \begin{subequations} \label{gen} 
            \begin{eqalignno}  
                  \Delta N & = (\sigma-\delta) N,  \\ 
                  \Delta D & = -\left(4 \mu + \frac{2 \sigma_1}{N} +  
\delta_2 \phi \right) D + 2 \mu L,\\
                  \Delta d & = -2 \mu (d- \frac{L}{2}). 
            \end{eqalignno} 
      \end{subequations} 
Here $\sigma=\sigma_1+\sigma_2$ and $\delta=\delta_1+\delta_2$ are the 
overall rates of speciation and extinction (per unit time interval per 
lineage), and $\phi$ is the proportion of the clade represented by a 
$T$-subclade that goes extinct. In general, these rates can change in time 
and/or with the clade size. With fossil data, in practice, it will generally 
not be possible to distinguish the two types of speciation events, and it 
will be very difficult to distinguish anagenetic from cladogenetic 
morphological 
change, but the overall rates of extinction and origination can be estimated.
Parameter $\mu=\mu_1+\sigma_2\mu_2$ is the overall probability of a morphological change (per trait per unit time interval per lineage) which incorporates both 
morphological and taxomonic rates of evolution. Below I describe a simple 
method for estimating $\mu$ from fossil data. 
 
The system of difference equations (\ref{gen}) describing the joint dynamics 
of $N, d$ and $D$, can be easily solved numerically. Below I consider several 
specific cases where solutions can be found analytically. There are also  
some general qualitative features of the dynamics of morphological evolution
which can be deduced from the form of equations (\ref{gen}). 
The right-hand side of equation (2b) has three negative 
and one positive terms. The latter, which is twice the expected number  
$\mu L$ of new traits per unit time interval per lineage, gives the maximum  
possible rate of increase in morphological disparity $D$.  
Each of the three negative terms is proportional to $D$ meaning they 
are negligible initially when the clade is confined to a small volume on the 
hypercube (when $D$ is small) but become increasingly important as the clade diversifies morphologically (when $D$ becomes larger). 
The first negative term in the right-hand side of equation (2b) is related 
to the geometric structure of the morphospace: 
as the clade expands in the morphospace, it becomes less and less probable 
that a random morphological change will lead outside the volume of the  
morphospace already occupied by the clade.  
The second term describes the reduction in $D$ because of the splitting of lineages into independent  
units without immediate and significant morphological changes. 
The third term specifies the reduction in $D$ due to the extinction of subclades.  Thus, equation (2b) predicts rapid initial increase in $D$ 
with a slowdown afterwards coming because of the geometric 
structure of the morphospace and the effects of extinction and speciation on 
morphological disparity.  The slowdown is expected even when all underlying
processes are time-homogeneous.

Let us turn now to variable $d$.
The general solution  of equation (2c) can be approximated as 
      \begin{equation} \label{d}
            d=\frac{L}{2}\left(1-exp( - \int_0^t 2 \mu dt) \right). 
      \end{equation} 
This shows that provided $\mu>0$, the average distance $d$ of the clade from
its ancestral state is always expected to increase monotonically towards the 
asymptotic value $L/2$. In particular, this dynamical feature is not affected
by changes in the rates of extinction, origination and morphological changes.

To illustrate the dynamics of the clade diversification let us assume  
that all parameters of the model are constant and that the rate of origination 
exceeds that of extinction. Let $R=\sigma-\delta$ be the clade's growth rate.
In this case, the clade size increases exponentially: 
      \begin{subequations} \label{exp} 
            \begin{equation}  
                  N(t)= exp(\tau), 
            \end{equation} 
the average pairwise distance within the clade tends to 
$L/(2+\delta_2 \phi/2\mu)$: 
            \begin{equation}   
                  D(t)= L e^{-a \tau} exp( be^{-\tau}) c b^a \left[   
                     \Gamma(-a,b e^{-\tau})-\Gamma(-a,b) \right], 
            \end{equation}  
whereas the average distance from the founder approaches $L/2$: 
            \begin{equation}   
                  d(t)=\frac{L}{2}\left[1-exp(-c\tau) \right]. 
            \end{equation}
      \end{subequations} 
\begin{figure}[tbh] 
\begin{center} 
\scalebox{0.35}{\includegraphics[1.5in,1.5in][8in,10.in]{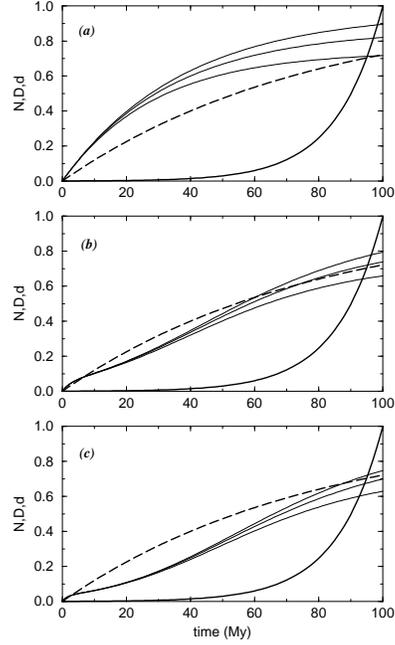}} 
\end{center} 
\caption{ {\small Dynamics of diversification as predicted by equations (4)
with $\sigma=0.32, \delta=0.25, \mu=0.006$ and the unit time interval
corresponding  
to one million years (cf. Foote 1996a). 
(a) Zero probability of speciation of the first type  
($\sigma_1=0, \sigma_2=0.32$),  
(b) Equal probabilities of speciation of the first and second type  
($\sigma_1=0.16, \sigma_2=0.16$),  
(c) Zero probability of speciation of the second type 
($\sigma_1=0.32, \sigma_2=0$).  
Thick lines represent clade size $N$ relative to the size achieved by  
the end of the time interval studied, $N_{max}=1096$. 
Dashed lines represent the average distance of the clade from the founder 
relative to the asymptotic value $L/2$.  
Thin lines show disparity $D$ (normalized relative to $L/2$) corresponding  
to the extinction every 10-th time step of a sub-clade representing  
10\% (top lines), 20\% (middle lines) and 30\% (lower lines) of the clade.}} 
\end{figure} 

Here $\tau=Rt, a=(4\mu+\delta_2 \phi)/R,b=2\sigma_1/R, c=2\mu/R$ and  
$\Gamma(x,y)$ is the incomplete gamma-function (e.g. Gradshteyn \&  
Ryzhik, 1994). The dynamics effectively  
depend on only three parameters, $\mu/R, \delta_2 \phi/R$ and $\sigma_1/R$,  
characterizing the probabilities of a morphological change, subclade
extinction 
and speciation of the first type relative to the clade's growth rate. 
Figure 1 illustrates the patterns of diversification predicted by equations 
(\ref{exp}). 
The numerical values used for the overall extinction and origination rates 
$\delta$ and $\sigma$ are the same as in Foote (1996a) where the estimate of  
$\delta$ was based on Raup's (1991) data whereas the origination rate $\sigma$
was set to produce an increase in diversity to about $1000$ lineages in  
100 million years. The numerical value used for $\mu$ was estimated from the 
blastozoan data (see below). Depending on parameter value, disparity $D$ 
can increase faster or slower than $d$. 
Note that even in the case of exponential increase in the clade size, 
the clade's disparity $D$ will approach $L/2$ asymptotically only  
if there is no sub-clade extinction ($\delta_2=0$).  
In other situations, the asymptotic value of $D$ will be smaller than  
$L/2$. For instance, if on average every 50-th time step a subclade  
representing 50\% of the clade goes extinct, then $\delta_2= 1/50 \times
1/2=.01, \phi=1/2$, and if $\mu=0.0025$, then $D$ will approach $L/3$. 
The Appendix lists several other specific cases where  
equations (2) can be solved analytically. In general, if the clade increases 
in size (if $R>0$), its morphological diversification is the fastest  
initially and slows down afterwards. The dynamics of $D$ are close to that  
in the exponential case, especially during initial stages.  
If the clade decreases in size (if $R<0$), its loss of morphological 
disparity is delayed relative to the loss in the number of lineages (cf. Foote
1993, 1996a).

\section{3. APPLICATION OF MODEL}

The model presented above provides a framework for studying the complex 
processes of clade diversification. It can be used to train our intuition 
about these processes, to identify key components, and to suggest hypotheses 
that can be tested against fossil data.
Can one use the observed dynamics of $D$ and $d$ for making quantitative 
inferences about the underlying processes? Using disparity $D$ is 
complicated  because both its dynamics and asymptotic value depend on a 
number of parameters that may change in time and be 
difficult to extract from fossil data, such as the relative importance of 
clade - vs. species - extinction, the relative importance of phyletic vs. 
cladogenetic change, and the proportion of speciation events that involve 
significant morphological change. In a wide variety of curcumstances, however,
distance $d$ is expected to increase regularly towards a fixed asypmtotic
value $L/2$. Equation (\ref{d}) implies that $-\ln(1-2d/L)= \int_0^t 2 \mu dt$. 
If the overall probability of a morphological change $\mu$ does not change  
(significantly) in time, the integral is simply $2\mu t$ and the dependence of
$q=1-2d/L$ on time should be a linear function on the semilog-scale. 
(In statistical physics variable $q$ is known as the average 
``overlap'' between binary sequences, e.g. Derrida and Peliti 1991.) This 
provides a simple test for approximate constancy of $\mu$. The constancy of
$\mu$ during a certain period would suggest the constancy of evolutionary 
rates and mechanisms.
Thus, if the rate of increase of disparity $D$ declines over time while
the rate of increase of $-ln(q)$ remains approximately constant, then
the explanation of the deceleration of morphological diversification
as a consequence of a change in evolutionary rates should be rejected. 
If $-\ln(q)$ changes as a linear function of time, the slope of the  
regression line gives an estimate for $2\mu$.  
In our model, the overall probability of a morphological change $\mu$ is
a sum of the phyletic component $\mu_1$ and the cladogenetic component $\mu_2\sigma_2$. Given some information about the patterns of origination
in a clade, it should be possible to help constrain the relative importance 
of phyletic and cladogenetic change. For example, a reduction in 
$\sigma_2$ is supposed to translate into a comparable reduction in $\mu$ unless 
the phyletic component is much larger than the cladogenetic component. 
Thus, if there is evidence for a change in speciation rate ($\sigma$)
without a proportional change in the rate of morphological evolution ($\mu$), this suggests a greater role for phyletic evolution than would be suggested by 
a concordant changes in $\mu$ and $\sigma$. 
The methods of extracting various information about $\mu$ from the clade level
data proposed above are potentially useful in general, but especially in 
those cases in which the phylogenetic information
needed to measure ancestor-descendant differences is unavailable.

\section{4. DIVERSIFICATION OF BLASTOZOANS} 
 
In light of the foregoing discussion, I use the model described above to
reanalyze the data on the 
morphological diversification of blastozoans (Foote 1992, 1996a).
These data provide one of the best illustrations 
of the pattern of accelerated early morphological diversification.  
The data represent sixty-five discrete characters measured for 147 species 
spanning across 12 stratigraphic levels from the Lower Cambrian to the Permian.
Morphological distances between species were measured as the total
number of differences divided  by the number of characters compared 
corresponding to $d/L$ and $D/L$ in the notation of the model. 
The time scale used is from Tucker \& McKerrow (1995) (with the exception
of the Carboniferous and Permian which are not covered by Tucker \& McKerrow
and for which I used the Harland {\em et al}. 1989 scale).
The data points were placed in the middle of the intervals. 
Morphological disparity ($D$) increases more rapidly than taxonomic 
diversity ($N$) reaching one half of the maximum observed level by the Late  
Cambrian and the maximum observed level by the Middle Ordovician (Figure 2a,b).
As emphasized earlier (Foote 1992, 1996a), during the period of this clade's expansion the rate of increase of $D$ apparently declines in time (Fig.2b). 
The taxonomic diversity $N$ grows through the Caradocian and decays after it suggesting a major change in the pattern of origination and extinction  
somewhere near the Caradoc-Ashgill boundary. 
This change coinsides with apparent drops  
in both $d$ and $D$. A drop in morphological disparity
$D$ can be caused by an increase in subclade extinction rates, an increase in
the rate of speciation of the first type, the selective extinction of
morphologically ``peripheral'' (relative to the founder) lineages, 
or the selective proliferation of morhologically ``central'' lineages, 
among other factors.  A drop in $d$ can be caused by extinction of 
morphologically ``peripheral'' lineages and/or by intensive speciation 
of ``central'' lineages. The overall decrease in taxonomic diversity  
$N$ between the Caradocian and Ashgillian suggests the possibility that it
may have been  increased extinction  
of morphologically peripheral lineages that caused  
the drop. Note that the decrease in morphological disparity is delayed  
relative to the decrease in taxonomic diversity.  
Distance $d$ continues to increase after the Ordovician while $D$ declines.  
The fact that $D$ is low and $d$ is high later in the clade history 
means that the clade forms a compact group evolving far away from the 
founder. (In this case it is the Blastoidea [Foote, personal communication].) 
This does not necessarily imply directionality in the processes governing 
clade evolution or species selection. Such behavior is expected for  
a completely random walk in a multidimensional space (cf. Charlesworth 1984; 
Bookstein 1987). 

As indicated by the apparent linearity on the semilog-scale (Figure 2c),
the dynamics of $d$ appear to be time-homogeneous from the Lower Cambrian 
through the Middle Ordovician and from the Upper Ordovician through the Upper Carboniferous 
with the drops in $d$ near the Middle Ordovician-Upper Ordovician boundary and in the Permian.
I used this as a justification for splitting the data set into two parts 
(for computing separate regression lines) and excluding the
Permian point. 
The estimates of $\mu$ for the periods from the Lower Cambrian through the 
Middle Ordovician and from the Upper Ordovician to Upper Carboniferous are $(5.8\mp 
0.4)\times 10^{-3}$ and $(3.6\mp 0.2) \times 10^{-3}$, respectively. 
For exponential processes, the time scale is usually characterized in terms 
of a half-life $T_{1/2}$. The half-life for $d$ is $\ln(2)/(2\mu)$.   
With $\mu=0.0058$ and $\mu=0.0036$, $T_{1/2}$ is about 60My and 96My,  
respectively. The linear regressions provide an excelent fit (the 
$r^2$ coefficients are $0.983$ and $0.986$), 
the slopes are significantly different from zero ($P<0.001$) 
and from each other ($P<0.01$). The increase 
in the quality of fit gained by fitting two separate lines rather than just
one line is significant at $P<.01$ (the Snedecor test). 
The numerical values reported above should be taken with some degree of
caution for some of the assumptions underlying the regression methods might 
be violated. The regression estimates are obviously sensitive to details of 
time scale and resolution.  
The appearance of more precise stratigraphic data would probably require
to reevaluate parameter estimates but is not expected to change our
qualitative conclusions about the diversification of blastozoans.

 \begin{figure}[tbh] 
 
\begin{center} 
\scalebox{0.35}{\includegraphics[1.5in,1.5in][8in,10.in]{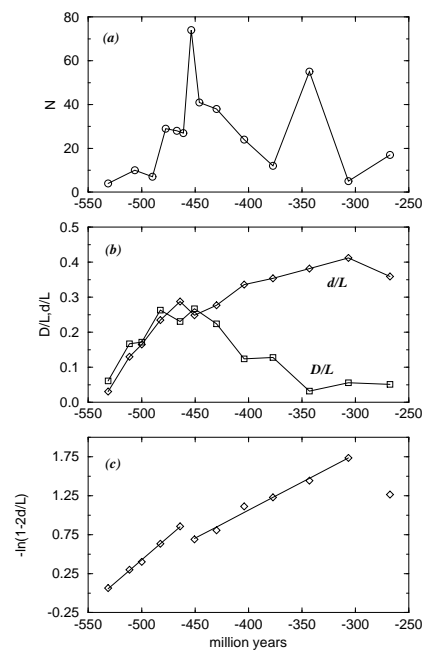}} 
\end{center} 
\caption{ {\small Reanalysis of the data (Foote 1992, 1996a) on  
the morphological and taxonomic diversification of Blastozoans. 
(a) The number of genera (which is assumed to be a reasonable proxy of 
the number of species $N$).  
(b) Morphological disparity and the average distance from the founder. 
For each stratigraphic level except the first one, the average  
distance from the species-founder $d$ was approximated by the average  
distance between the current level and the first level. For the first  
stratigraphic level, $d$ was approximated as half of $D$.  
(c) Transformed values of $d$ and the corresponding linear regression 
(see text for more details). 
In Figure (a) the 14 data points correspond to the following stratigraphic
levels: Cambrian (Lower, Middle/Upper), Ordovician (Tremadoc, Arenig, 
Llanvim, Llandeilo, Caradoc, Ashgill), Silurian, Devonian
(Lower, Middle/Upper), Carboniferous (Lower, Upper) and Permian.
Figures (b) and (c) use a coarser resolution with only 3 Ordovician
intervals (Lower, Middle and Upper).}} 
\end{figure} 

Overall, the data are compatible with a moderate 
($38\%$) reduction in $\mu$ that took place near the boundary of the Middle
Ordovician and the Upper Ordovician. 
A decrease in the rate of morphological evolution was also 
advocated on the basis of the shape of the disparity curve (Foote 1992)
and on estimates of morphological separation between closely related 
taxa (Wagner 1995a). 
The reduction in $\mu$, however, is not responsible for the apparent
deceleration of morphological diversification observed during the first 
third of the clade's history when $\mu$ was apparently constant.

The fossil record shows that there has been significant 
decline in rates of origination within major taxa through their histories 
(e.g. Van Valen 1985; Sepkoski 1998). 
The moderate size of the decrease in $\mu$ observed for blastozoans together 
with a significant decrease in speciation rates would suggest that for
this clade morphological evolution is driven mainly by anagenetic rather than
cladogenetic changes. Although the fact that blastozoan taxonomic 
diversity increased initially and declined later on (see Fig.2a) is
compatible with the decline in speciation rates this is not a definite
conclusion. The decline in taxonomic diversity can be caused by
an increase in extinction rates rather than by a decrease in origination 
rates. Additional data are needed for reaching more precise conclusions.

\section{5. DISCUSSION}

It is important to realize that apparent secular changes in the rates of
morphological evolution for a clade as a whole do not necessarily mean  
secular changes in the processes acting at the level of individual lineages. 
In particular, the observed deceleration of morphological disparity does 
not necessarily imply a decline in the size or probability of morphological
changes  
for individual lineages. The model presented here has demonstrated that such a deceleration 
is expected from the geometric structure of the morphospace and the
effects  
of extinction and speciation on morphological disparity even when all  
relevant processes are time-homogeneous. These theoretical  
predictions appear to be very robust. In particular, the differences  
between exponential and logistic growth in the taxonomic diversity do not 
translate into significant changes in the corresponding dynamics of
morphological diversification.  Our basic conclusions will definitely be
valid if the traits have more than two discrete states and should be valid 
if the morphological space is continuous rather than discrete 
as long as it is finite. In the continuous case, the effects of speciation  
of the first type and sub-clade extinction on disparity will be similar 
to that in the discrete model considered here. 
Although demonstrating the existence, nature, and importance of morphological
boundaries can be difficult (McShea 1994, Foote 1996a),
it is intuitively obvious that, given they exist, the
process of divergence will slow down even if these two factors are absent. 
The potential importance of the model I present is not only that it 
quantifies and
trains our intuitions, but also that it allows one to test whether, in the
case of discrete characters, the observed deceleration in morphological
diversification is likely to be a simple consequence of the nature of
evolution on a binary hypercube, and thus whether it is necessary to invoke temporal heterogeneities in evolutionary rates and mechanisms to explain 
an observed pattern. The model makes  
falsifiable predictions about the dynamics of morphological disparity and  
the average distance of the clade from its ancestral state, provides a 
simple method to evaluate the rate of morphological evolution and
suggests an approach for comparing the importance of anagenetic and
cladogenetic changes in morphological diversification. In the case of
blastozoans, I find no evidence that major changes in evolutionary rates
and mechanisms are responsible for the deceleration of morphological
diversification seen during the period of this clade's expansion.
At the same time, there is evidence for a moderate decline in overall rates of
morphological diversification concordant with a major change (from
positive to negative values) in the clade's growth rate.
 
The model has its limitations. The most significant is probably that 
it describes only average behavior and says nothing about variation which  
will always  be present in the fossil record (and numerical simulations). 
In particular, this makes it difficult to evaluate the statistical power  
of the test of time-homogeneity proposed above. 
The model developed above makes no restrictions on morphology in the 
sense that all character 
combinations are assumed to be potentially realizable. In terms of the 
metaphor of ``adaptive landscapes'' (Wright 1932), the model assumes a  
``flat'' landscape similar to those in 
models of neutral molecular evolution (e.g. Derrida \& Peliti 1991). 
In general, because of genetic, developmental, or ecological constraints  
some of the possible character combinations can be prohibited. In this 
case, the morphospace will be mathematically equivalent to a hypercube with 
``holes'' (with ``holes'' representing prohibited character combinations)  
and the corresponding adaptive landscape will be  
``holey'' (Gavrilets 1997, 1999; Gavrilets \& Gravner 1997; Gavrilets
{\em et al.} 1998) rather than ``flat''. 
If the proportion of holes is not extremely high, ``viable'' character 
combinations will form a ``giant'' cluster extending through the whole  
morphospace. A characteristic signature of 
a random walk on the giant cluster appears to be a stretched exponential
dependence 
of overlap $q$ on time (e.g. Lemke \& Campbell 1996): 
$q(t) \sim exp(-(t/\tau)^{\beta})$, where $\tau$ and $\beta\leq 1$ are 
parameters (with no holes $\beta=1$). 
The fitting of the stretched exponential curve to blastozoan data 
has led to inconclusive results: although the fit is good, it is not better
than the fit of a simple exponential curve described above. More detailed 
data sets are needed for more precise conclusions. 
 
\begin{center} 
Acknowledgments.  
\end{center} 
I are grateful to Mike McKinney for helpful discussions and suggestions
and to Mike Foote for the data used, advice, critique, and suggestions.
Supported by National Institutes of Health grant GM56693.\\

%\newpage 
 
\section{Literature} 
 
Bookstein, F. L. 1987 Random walk and the existence of 
evolutionary rates. {\em Paleobiology} {\bf 13}: 446-464. 
 
  \label{Brig92} Briggs, D. E. G., Fortey, R. A. \& Wills, M. A. 1992 
Morphological disparity in the Cambrian. {\em Science} {\bf 256}: 1670-1673. 
 
  Charlesworth, B. 1984 Some quantitative methods for studying 
evolutionary patterns in single characters. {\em Paleobiology} {\bf 10}:
308-318. 
 
  \label{Crow70} Crow, J. F. \& Kimura, M. 1970 {\em Introduction to 
population genetics theory}. New York: Harper and Row. 
 
  \label{Derr91} Derrida, B. \& Peliti, L. 1991 Evolution in a flat 
landscape. {\em Bull. Math. Biol.} {\bf 53}: 255-382.

  \label{ElGo72}Eldredge, N. \& Gould, S. J. 1972 Punctuated equilibria: 
an alternative to phyletic gradualism. In  {\em Models in 
paleobiology} (ed. Schopf, T. J . M.), pp. 82-115. San Francisco: 
Freeman, Cooper and Co.  
 
  \label{Erwi94} Erwin, D. H. 1994 Early introduction of major 
morphological innovations. {\em Acta Palaentol. Polonica} {\bf 38}: 281-294. 
 
  \label{Erwi87} Erwin, D. H., Valentine, J. W. \& Sepkoski, J. J.  
Jr. 1987 
A comparative study of diversification events: the early Paleozoic versus 
the Mesozoic. {\em Evolution} {\bf 41}: 1177-1186. 

  \label{Foote91} Foote, M. 1991 Morphological and taxonomic diversity  
in a clade's history: the blastoid record and stochastic simulations.  
{\em Contributions from the Museum of Paleontology, University of Michigan}  
{\bf 28}: 101-140. 
 
  \label{Foote92} Foote, M. 1992 Paleozoic record of morphological  
diversity in blastozoan echinoderms. {\em Proc. Natl. Acad. Sci. USA} 
{\bf 89}: 7325-7329. 
 
  Foote, M. 1993 Discordance and concordance between morphological and
taxonomic diversity. {\em Paleobiology} {\bf 19}:185-204. 

  \label{Foote94b} Foote, M. 1994 Morphology of Ordovician-Devonian 
crinoids. {\em Contributions from the Museum of Paleontology, University  
of Michigan} {\bf 29}: 1-39. 
 
  \label{Foote95a} Foote, M. 1995 Morphology of Carboniferous and 
Permian crinoids. {\em Contributions from the Museum of Paleontology,  
University of Michigan} {\bf 29}: 135-184. 
 
  \label{Foote96a} Foote, M. 1996a Models of morphological  
diversification. In {\em Evolutionary Paleobiology} (eds. Jablonski, D.,
Erwin, D. H. \& 
Lipps, J. H.) pp. 62-86. Chicago: The University of Chicago Press. 
 
  Foote, M. 1996b Ecological controls on the evolutionary recovery of post-
Paleozoic crinoids. {\em Science} {\bf 274}: 1492-1495.

   \label{Gavr97} Gavrilets, S. 1997 Evolution and speciation on holey 
adaptive landscapes. {\em Trends Ecol. Evol.} {\bf 12}: 307-312. 
 
  \label{Gavr99} Gavrilets, S. 1999 A dynamical theory of speciation  
on holey adaptive landscapes. {\em Amer. Natur.} (submitted) 
 
  \label{Gavr98} Gavrilets, S., Li, H. \& Vose, M. D.  1998 Rapid  
parapatric speciation on holey adaptive landscapes. {\em Proc. R. Soc. Lond. 
B.} 
 
  \label{GaGr} Gavrilets, S. \& Gravner, J. 1997 Percolation on the  
fitness hypercube and the evolution of reproductive isolation. {\em J.  
Theor. Biol.} {\bf 184}: 51-64. 
 
    \label{GrRy94} Gradshteyn, I. S. \& Ryzhik, I. M. 1994 {\em Tables of 
Integrals, Series, and Products}. Fifth Edition. San Diego: Academic Press.

  \label{Kauf93} Kauffman, S. A. 1993 {\em The origins of order}.  
Oxford: Oxford University Press. 
 
  \label{Mayr42}  Mayr, E. 1942 {\em Systematics and the origin of 
species}. New York: Columbia University Press. 
 
  \label{Mayr63}  Mayr, E. 1963 {\em Animal Species and Evolution},  
Cambridge: Harvard University Press. 

McKinney, M. L. 1990. Classifying and analyzing evolutionary trends.
In {\em Evolutionary Trends} (ed. McNamara, K. J.) pp. 75-118. Tucson:
University of Arizona Press.

McKinney, M. L. 1997. Extinction vulnerability and selectivity: combining
ecological and paleontological views. {\em Ann, Rev. Ecol. Syst.} {\bf 28}:
495-516.
 
  \label{McSh93} McShea, D. W. 1993 Arguments, tests and the Burgess 
Shale - a commentary on the debate. {\em Paleobiology} {\bf 19}: 399-402. 

   McShea, D. W. 1994 Mechanisms of large-scale evolutionary trends {\em
Evolution} {\bf 48}: 1747-1763.
 
  \label{Mill96} Miller, A. I. \& Foote, M. 1996 Calibrating the Ordovician 
radiation of marine life: implications for Phanerozoic diversity trends. 
{\em Paleobiology} {\bf 22}: 304-309. 
 
  \label{Lemke96} Lemke, N. \& Campbell, I. A. 1996 Random walks in a 
closed space. {\em Physica A} {\bf 230}: 554-562. 
 
  \label{Lync89} Lynch, J. D. 1989 The gauge of speciation: on the 
frequencies of different modes of speciation. In  
{\em Speciation and its consequences} (eds. Otte, D. \& Endler, J.) 
pp. 527-553. Sunderland, MA: Sinauer Associates. 
 
  \label{Raup73} Raup, D. M., Gould S. J., Schopf, T. J. M. \& 
Simberloff, D. S.  1973 Stochastic models of phylogeny 
and the evolution of diversity. {\em J. Geology} {\bf 81}: 525-542.  
 
  \label{Raup74} Raup, D. M. \&  Gould, S. J. 1974 Stochastic simulations 
and the evolution of morphology: towards a nomothetic paleontology.  
{\em Systematic Zoology} {\bf 23}: 305-322.

  Raup, D. M. 1991 A kill curve for Phanerozoic marine species. 
{\em Paleobiology} {\bf 17}: 37-48. 
 
  Sepkoski, J. J. Jr. 1998 Rates of speciation in the fossil record. 
{\em Phil. Trans. R. Soc. Lond. B} {\bf 353}: 315-326. 
 
  \label{Slat81} Slatkin, M. 1981 A diffusion model of species selection. 
{\em Paleobiology} {\bf 7}: 421-425.

\label{TuMc95} Tucker, R. D. and W. S. McKerrow. 1995 Early Paleozoic
chronology: a review in light of new U-Pb zircon ages from Newfoundland and
Britain. {\em Can. J. Earth Sci.} {\bf 32}: 368-379.

  \label{Val69} Valentine, J. W. 1969 Patterns of taxonomic and  
ecological structure of the shelf of benthos during Phanerozoic time.  
{\em Palaeontology} {\bf 12}: 684-709. 
 
  \label{Val80} Valentine, J. W. 1980 Determinants of diversity in 
higher taxonomic categories. {\em Paleobiology} {\bf 6}: 444-450. 
 
  \label{Val94} Valentine, J. W., Collins, A. G. \& Meyer, C. P. 1994 
Morphological complexity increase in metazoans. {\em Paleobiology} {\bf 20} 
: 131-142. 
 
  Van Valen, L. 1985 Why and how do mammals evolve unusually rapidly? 
{\em Evolutionary Theory} {\bf 7}: 127-132. 
  
  \label{Wagn95a} Wagner, P. J. 1995a Systematics and the fossil record. 
{\em Palaios} {\bf 10}: 383-388. 
 
  \label{Wagn95b} Wagner, P. J. 1995b Testing evolutionary constraint  
hypothesis with early Paleozoic gastropods. {\em Paleobiology} {\bf 21}:  
248-272. 
 
  \label{Will94} Wills, M. A., Briggs, D. E. G. \& Fortey, R. A. 1994 
Disparity as an evolutionary index: a comparison between Cambrian and Recent 
arthropods. {\em Paleobiology} {\bf 20}: 93-130.

\subsection{Appendix} 
 
Each new trait in a lineage increases or decreases its distance  
from the founder by one with probability $1-d/L$ and $d/L$, respectively. 
Extinction of a lineage is not expected to change $d$.  
Thus, the overall expected change in $d$ is $\mu[1\times (1-d/L) -1 \times 
d/L]$ which reduces to (2c). Consider the dynamics of $D$.  
Because each lineage in a randomly chosen pair can evolve morphologically,  
the rate of change in $D$ induced by phyletic evolution and by cladogenesis  
is twice as big as in the case of $d$ and is $-4 \mu (D-L/2)$.  
Speciation events  
of the first type will decrease $D$ because the distance between the  
immediate descendants of a species that has split will be zero. 
Let $P$ be the probability that two randomly chosen lineages originated from 
a split of a species in the previous time interval. If $\overline{k}$ and  
$var(k)$ are the average and the variance of the number of ``offspring'' 
species that a species leaves (counting itself) in the next time interval, 
then probability $P$ can be represented as $P \approx  
(var(k)/\overline{k}+\overline{k}-1)/N$ (Crow \& Kimura 1970). Assuming  
that a proportion  
$\delta$ of species go extinct whereas the remaining species survive and  
speciate, $\overline{k}=1+\sigma-\delta, var(k) = \sigma+\delta$ where the  
last equality assumes that both extinction and speciation rates are small. 
Thus, $P \approx 2\sigma/N$ (cf. Derrida \& Peliti 1991). Distance $D$ after 
the speciation events of the  
first type can be represented as $(1-P \sigma_1/\sigma)\times D+  
P \sigma_1/\sigma \times 0$. Thus, the expected reduction in $D$ due to the  
speciation events of the first type is 
$-2\sigma_1/N$.  Extinction of individual lineages  
is not expected to change $D$. Let us assume that there are $f$  
$T$-subclades and let $D_b$ be the average morphological distance between two
$T$-subclades. The average distance within the whole clade can be approximated
as $(N/f)^2 D_b f(f-1)/N^2$. Accordingly, the average distance after  
extinction of a $T$-subclade is approximately $(N/f)^2 D_b (f-1)(f-2)
[(1-1/f)N]^2$. 
This shows that extinction of a $T$-subclade reduces $D$ by $D/f^2$. Let
$\eta$ 
be the probability of extinction of a $T$-subclade per 
unit time interval and 
$\phi=1/f$ be the proportion of the clade that goes extinct. Then the rate 
of reduction in $D$ due to subclade extinction is approximately $-\delta_2
\phi$ 
where $\delta_2=\eta \phi$. Thus, the overall change in $D$ is given by  
equation (2b).

In the main text, a case with all parameters constant was considered. 
Here I list several other cases that can be treated analytically. 
Let the growth rate decrease linearly with the clade size: 
$R\equiv \sigma-\delta=r(1-N/K)$. Then the clade size approaches the  
``carrying capacity'' $K$ according to the logistic curve $ 
N(t) = K e^{\tau}/(e^{\tau}+K-1)$ where $\tau=rt$. 
The difference between exponential and logistic growth in $N$ should be 
most important after a transient time when the exponential model predicts
very large values of $N$ whereas in the logistic model $N$ approaches the
carrying capacity $K$. However if $N$ is large, the second term in the 
right-hand side of equation (2b) is negligible. Thus, the difference between exponential and logistic models for $N$ is not expected to translate into 
significant changes in the dynamics of morphological evolution if $K$ is 
not too small. Below I make this argument more precise.
The decrease in the growth rate $R$ with the clade size can result from  
decrease in the origination rates and/or increase in the extinction rates.  
I assume that other parameters do not change. 
If $\sigma_2=const$, then the dynamics of $d$ are still described by  
equation (4c). 
If $\sigma_2$ decreases linearly with the clade  
size $N$ from $\sigma_2(1)$ to $\sigma_2(K)$, then the dynamics of $d$ are  
approximated by equation 
      \begin{equation} \label{l2} 
            d(t)= \frac{L}{2} \left[ 1- e^{-C\tau}  
            \left( 1+\frac{e^{\tau}}{K} \right)^{2\mu_2 \Delta \sigma_2/r}
\right].  
      \end{equation} 
Here $C=2\mu^*/r, \mu^* = \mu_1+\mu_2 \sigma_2(1)$, $\Delta \sigma_2= 
\sigma_2(1)-\sigma_2(K)$ is the overall change in $\sigma_2$, 
and it is assumed that $K>>1$. 
To solve equation (2b) analytically when extinction and/or origination rates 
change with $N$ one needs additional simplifying assumptions. If $\delta_1$
changes 
with $N$ whereas all other rates are constant, the dynamics of $D$ are 
described by (4b) with $a=(4\mu+2\sigma_1/K+\delta_2 \phi)/r,\  
b=2(1-1/K)\sigma_1/r, 
c=2\mu/r$. Note that if the ``carrying capacity'' $K$ is large, coefficients 
$a,b$ and $c$ are close to the values corresponding to the exponential 
growth case and the dynamics of $D$ under exponential and logistic 
growth are similar. If $\sigma_1$ decreases linearly 
with the clade size $N$ whereas other rates are constant, the $D$  
evolves according to (3b) with  $a=(4\mu+2(\sigma_1(1)-r)/K)/r),  
b= 2(1-1/K)\sigma_1(1)/r,c=2\mu/r$. If the growth rate $r$ is small 
relative to the origination rate $\sigma_1(1)$, the dynamics of $D$ will be 
similar to that under exponential growth. If all speciation events are of 
the second type ($\sigma_1=0$) and there is no family extinction  
($\delta_2=0$), then the dynamics of $D$ are described by the  
right-hand side of equation (\ref{l2}) with $C=4\mu^*/r$. 
Let the clade size decrease linearly in time: $N(t)=N(0)-rt$. Then 
the average pairwise distance $D$ changes according to equation 
      \begin{eqalignno*}
            D(t)= & D(0) e^{-a\tau} \left( \frac{N(t)}{N(0)} \right)^b \\  
+  & c a^{b-1} e^{aN(t)} N(t)^b \left[ 
                        \Gamma(1-b,aN(t))-\Gamma(1-b,aN(0) \right], 
      \end{eqalignno*} 
where $\tau=rt$, parameters $a,b$ and $c$ are defined below equation (4b), 
and $D(0)$ is the initial value of $D$.

\end{document}